\documentclass[twocolumn, showpacs, preprintnumbers, amsmath, amssymb, superscriptaddress, pra]{revtex4}
\usepackage{graphicx}
\usepackage{subfigure}
\usepackage{dcolumn}
\usepackage{bm}
\usepackage{epsf}
\usepackage{amssymb}
\DeclareGraphicsExtensions{.eps}
\usepackage{epstopdf}
\usepackage{natbib}
\usepackage{graphicx}
\begin{document}

\author{David S. Simon}
\email[e-mail: ]{simond@bu.edu} \affiliation{Dept. of Physics and Astronomy, Stonehill College, 320 Washington Street, Easton, MA 02357}
\affiliation{Dept. of
Electrical and Computer Engineering \& Photonics Center, Boston University, 8 Saint Mary's St., Boston, MA 02215, USA}
\author{Casey A. Fitzpatrick}
\email[e-mail: ]{cfitz@bu.edu} \affiliation{Dept. of Electrical and Computer Engineering \& Photonics Center, Boston University, 8 Saint Mary's St., Boston, MA
02215, USA}
\author{Alexander V. Sergienko}
\email[e-mail: ]{alexserg@bu.edu}
\affiliation{Dept. of Electrical and Computer Engineering \& Photonics Center, Boston
University, 8 Saint Mary's St., Boston, MA 02215, USA}
\affiliation{Dept. of Physics, Boston University, 590 Commonwealth
Ave., Boston, MA 02215, USA}

\begin{abstract}
The concept of directionally-unbiased optical multiports is introduced, in which photons may reflect back out the input direction. A linear optical
implementation is described, and the simplest three-port version studied. Symmetry arguments demonstrate potential for unusual quantum information processing
applications. The devices impose group structures on two-photon entangled Bell states and act as universal Bell state processors to implement probabilistic
quantum gates acting on state symmetries. These multiports allow optical scattering experiments to be carried out on arbitrary undirected graphs via linear
optics, and raise the possibility of linear optical information processing using group structures formed by optical qudit states.
\end{abstract}

\title{Group Transformations and Entangled-State Quantum Gates\\ With Directionally-Unbiased Linear-Optical Multiports}

\pacs{42.50.Ex,42.50.-p,42.79.-e}

\maketitle

\section{Introduction}\label{introsection}

 Symmetry has long been a guiding principle in physics. Central to linear optics is the beam splitter (BS), which has reflection symmetry
as well as a time-reversal symmetry in which a photon may either enter or leave any of the four ports. However, the BS is asymmetric in another sense: once a
photon enters a port, it may not leave from the \emph{same} port: a photon in one port breaks the symmetry. We call this inability to reverse direction as
\emph{directional bias}. Here we introduce a linear optical arrangement that restores the symmetry: it is \emph{directionally unbiased}. This arrangement, which
can be thought of from multiple viewpoints (as a directionally unbiased multi-port BS, as a resonant cavity with three or more exit directions, or as a
scattering vertex for an undirected graph), is generalizable to any number of ports $n\ge 3$; we focus on the simplest case of $n=3$. Despite its simplicity, the
high degree of symmetry and the ability of photons to reverse direction leads to interesting properties that allow it to be used as the basis for a variety of
applications related to quantum information processing.

Finding practical means of transmitting higher-dimensional optical states (qudits) with high information capacity is a longstanding goal of quantum
communication, but here we raise a further possibility: increasing the number of \emph{operations} that can be applied to qudits, and arranging for these
operations to form a mathematical group. For example, instead of information processing using a one-dimensional string of $N$ orbital angular momentum states
connected in a chain by a single operator and its adjoint, it should be possible to use a set of $N$ states lying on some two- (or higher-) dimensional manifold,
and connected to each other by transformations of some $d$-dimensional group, so that $N\times d$ parameters are now needed to specify all possible transitions
between states. This would allow the power of group theory to be brought to bear on problems in quantum information processing and quantum communication to a
greater degree than it has previously, and could potentially offer increased speed and flexibility in areas such as quantum computation \cite{falci}. As one example of the new
possibilities, consider that group-based quantum cryptography protocols could be constructed in which the qudits being communicated may lie on points of a
collection of overlapping sets, and in which different sets form representations of different groups. Then, in order to extract information about the key,
an eavesdropper might need to obtain not just the \emph{qubits} being sent, but also knowledge of the abstract \emph{group} that joins them and of the group
\emph{representation} to which they belong. Group structures have already been shown to be useful for several different purposes in classical cryptography
\cite{blackburn,vasco}, but have not been widely applied in quantum cryptography protocols.

In the following we show, using only linear optics, that unbiased optical three-ports can encode Bell states \cite{genovese} in such a way that they form a representation of the
Klein Vierergruppe or 4-group. Generalized multiports and larger sets of states may allow implementations of larger group representations. We also apply the
multiport to an example of processing entangled-photon states, with bits encoded into the state's symmetry.
Generalizations and conclusions are then discussed. Elsewhere we will show that networks of unbiased multiports can produce quantum walks with unusual features.

\begin{figure}
\centering
\includegraphics[width=2.0in]{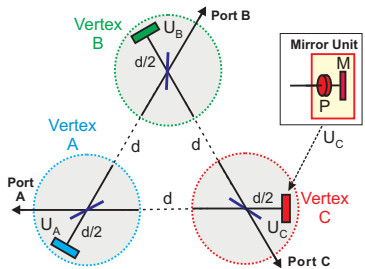}
\caption{(Color online.)  Schematic of the directionally-unbiased three-port. Beam splitters connect input ports to two internal edges
and a mirror unit $U$. $U$ is composed of a mirror $M$ and phase plate $P$ (inset).
Reflection at a mirror unit imparts to each photon a total phase factor of $-i$. }\label{trianglefig}\end{figure}

Quantum walks are of great interest for implementing quantum algorithms \cite{aharanov,kempe,andraca,portugal,childs}. In Refs. \cite{fh1,fh2,fh3}, quantum walks
on graphs were investigated, with scattering at the vertices. Particular attention was paid to an example with three-edge vertices that scatter both backward and
forward; in our terminology they are directionally unbiased. The initial motivation for the arrangement presented here was to implement such graphs using only
linear optics.

In the current paper, we do not yet make use of the full power of the arrangement to be presented. In particular, the vertices of the multiport have free
parameters (mirror reflectances and phase shifts) that can be varied to produce further effects; but in the present paper we keep all of these parameters fixed
and require them to have the same values at all vertices of the multiport. One effect of this is clear: by making the vertices identical we are restricting
ourselves to Abelian group structures. By allowing each vertex of the multiport to have different parameter values, it seems likely that non-Abelian structures
could be generated as well, although this remains to be investigated in detail.

\section{Directionally-Unbiased Optical Three-ports}\label{devicesection}  The arrangement shown in Fig. \ref{trianglefig} has three input/output ports, labeled $A$, $B$, and $C$, each
attached to two internal edges by a non-polarizing $50/50$ BS ($r=it={i\over \sqrt{2}}$). Light exiting the remaining port of each BS strikes a mirror normally.
Phase plates in front of each mirror add adjustable phase shifts. We keep the phase shift constant at $-{{3\pi}\over 4}$ per passage, so the total phase factor
gained at each mirror is $e^{-i\pi/2}=-i$ (including $-1$ from the mirror itself). This choice makes all exit amplitudes pure imaginary.  Beam splitter and
mirror losses are neglected.

Assume equal distances $d$ between beam splitters, with BS-to-mirror distance $d/2$. Then $T={d\over c}$ is the time between successive photon-BS encounters. The
probability of not exiting the device decreases exponentially to near zero within several steps (Table \ref{pathtable}), so for small enough $d$ (such that $T$
is small compared to the photon coherence time) the exit can be treated as instantaneous.

Each photon path through the system has amplitude of absolute value $2^{-N/2}$, after $N$ beam splitter encounters. The number of paths increases more slowly
(roughly linearly) with $N$, so output states can be calculated to a good approximation from the shortest few paths. For single-photon input at $A$, summing the
amplitudes of all possible paths of length $N$ is straightforward. Amplitudes up to $N=10$ are tabulated in Appendix A. For $N\ge 3$, the exit probability at the
$N$th step is $3/2^N$ for even $N$; the probabilities vanish odd $N$.

\begin{table}
\caption{Amplitudes for paths exiting at time $t=NT$, assuming input at $A$. Columns $2$-$4$ give amplitudes
for exit at each port. The last columns give exit probability at $t=NT$ and cumulative exit probability for $t\le NT$.}
\begin{tabular}{c|c|c|c|c|c} \hline $N$ & $A$ exit & $B$ exit  & $C$ exit  & Exit Prob. & Cumul. Exit Prob.\\ \hline
$2$ & $0$ & ${i\over 2}$ & ${i\over 2}$ & $1\over 2$ & $.5$ \\
$4$ & $-{i\over 2}$ & ${i\over 4}$ & ${i\over 4}$ & $3\over 8$ & $.875$ \\
$6$ & ${i\over 4}$ & $-{i\over 8}$ & $-{i\over 8}$ & $3\over {32}$ & $.96875$ \\
$8$ & $-{i\over 8}$ & ${i\over {16}}$ & ${i\over {16}}$ & $3\over {128}$ & $.99219$ \\
$10$ & ${i\over {16}}$ & $-{i\over {32}}$ & $-{i\over {32}}$ & $3\over {512}$ & $.99805$ \\  \hline \end{tabular}\label{pathtable}\end{table}

Ignoring overall normalization, the resulting state is $|\psi (t)\rangle = a(t)|A\rangle +b(t)\left( |B\rangle +|C\rangle \right) + |\psi_{int}
(t)\rangle $, where
\begin{small}\begin{eqnarray} a(t)& = & \left( - {i\over 2}\Theta (t-4T)+ {i\over 4}\Theta (t-6T)\right. \nonumber\\
& & \quad \left. -{i\over 8}\Theta (t-8T) +{i\over {16}}\Theta (t-10T) \right)  + {\cal O}\left( 2^{-5} \right)\\
b(t) &=& \left( {i\over 2}\Theta (t-2T) +{i\over 4} \Theta (t-4T) -{i\over 8} \Theta (t-6T) \right. \nonumber\\ & & \quad\left. + {i\over {16}}\Theta (t-8T)
-{i\over {32}} \Theta (t-10T)\right)+ {\cal O}\left( 2^{-6} \right).
\end{eqnarray}\end{small} $|A\rangle$, $|B\rangle$, and $|C\rangle$ are shorthand for single photon states
$|1\rangle_A|0\rangle_B|0\rangle_C$, $|0\rangle_A|1\rangle_B|0\rangle_C$, and $|0\rangle_A|0\rangle_B|1\rangle_C$. Step function $\Theta (t)$
vanishes for $t<0$ and equals unity for $t>0$. Finally, $|\psi_{int}(t)\rangle$ is the state with the photon still inside the device; its exponentially
decaying amplitude can also be obtained by summation of paths.

The transient internal state $|\psi_{int}(t)\rangle$ decays with characteristic time taken conservatively as $T_c\approx 10T$. If detector response time $T_D$
exceeds $T_c$  the possible photon paths are indistinguishable, so the exit state is a coherent superposition of outputs at different ports. Integrated onto a
chip, $d$ could be very small; as a conservative example take $d=.1\; mm$. Then $T_c=3.3 \; ps$, allowing sampling at rates up to $.3\; THz$.  Using a table-top
setup, $d$ would be on the order of centimeters, with proportionally lower maximum sampling rate.

For information processing, timing information must be supplied to synchronize gates. For instance, a CW source can be gated to form pulses. Assuming a Gaussian
temporal envelope, the pulse duration $\Delta t$ and spectral width $\Delta \nu$ are related by $\Delta t\Delta \nu ={1\over {4\pi}}$\; \cite{saleh}. For
example, pulses of $\Delta t\sim 10^{-10}\; s $ have $\Delta\nu\sim 1\; GHz$, giving coherence time and length $\tau_{coh}={1\over {\Delta\nu}}\approx 1\; ns$
and $l_{coh}\approx 30\; cm$, consistent with the constraints above. Using parametric down conversion is another possibility, with signal sent into the
multiport, and idler heralding the event to provide timing information. In this case, signal coherence times are of order $\tau_{coh}\sim 1\; ps$, leading to
coherence length $l_{coh}\sim 10^{-4}m$, bordering on the acceptable limit in the example above.

If the unit is constructed on a chip, then the refractive index of the chip will increase the transit time $T$ (by a factor of $2$--$4$ for materials like
silicon nitride and silicon). This will tighten the constraints somewhat. Using the numbers above, this would rule out parametric down conversion, but it still
leaves the gated CW source as a plausible possibility, with a correspondingly reduced sampling rate.

Before treating the three-port quantitatively, let us look at some potential problems that could arise.
First, when  $T_D<T_c$, different exit times for the same input become distinguishable. In this case, the output will be a mixed state, and the system behaves
classically. Further, when the wavepacket has finite coherence time, the corresponding frequency spread in the packet will cause different parts of the packet to go out of phase with each other, introducing a phase spread of up to $\Delta \phi =d\; \Delta k = {{2\pi d}\over {c\tau_{coh}}}$ during propagation along each edge. This will affect the shape of the wavepacket through interference, as well as introducing some spectral dependence into the output. All of these troublesome effects are to be avoided for high quality output;  so to keep their effects at a minimal level is is necessary to make sure that the inequalities $\tau_{coh} >>T_D>T_c$ hold.

A further potential problem is that when the multiport is used as a vertex in a quantum walk, the number of steps over which the walk will retain its quantum nature will be limited because of the broadening of wavepackets that results from the existence of different-length paths through the multiport. The maximum number of steps possible for the quantum walk will then be given roughly by the ratio ${{\tau_{coh}}/{T_c}}$; this again highlights the importance of maintaining the inequalities given above. By the same reasoning, when the multiport is used as a quantum gate the number of such gates that can be connected in sequence will be limited by the same ratio. For the examples given above, it can be seen that the gated laser pulse should be able under ideal conditions to maintain a quantum walk for dozens of steps, while the down conversion example will not be able to support walks of more than a couple of steps.

Given the considerations above, we now assume that $\tau_{coh} >>T_D>T_c$. Then the input $|\psi_{in}\rangle =|A\rangle$ leads to
\begin{eqnarray} |\psi_{out}\rangle &\approx & -i\left( {1\over 2}-{1\over {4}}+{1\over 8}-{1\over {16}}\right) |A\rangle \\ & & \quad
+ i\left( {1\over 2}+{1\over {4}}-{1\over 8}+{1\over {16}}-{1\over {32}}\right)
\left( |B\rangle + |C\rangle \right)\nonumber
\end{eqnarray}

The $A\to A$ part gives the first few terms of a geometric series. Assuming this pattern continues, the $A\to A$ transition amplitude is therefore
\begin{equation}-{i\over 2}\sum_{n=0}^\infty \left( -{1\over 2}\right)^n = -{i\over 2} \left( {1\over {1+1/2}} \right) =-{i\over 3}. \end{equation} By similar extrapolation, the $A\to
B$ and $A\to C$ coefficients are both
\begin{equation} i\left[ {1\over 2} + {1\over 4} \sum_{n=0}^\infty \left( -{1\over 2}\right)^n \right]  =i \left( {1\over 2}+{1\over 6} \right)
={2\over 3}i .\end{equation}

Rotational symmetry allows transition amplitudes for input at other ports to be obtained by cyclic permutation, giving the long-time single-photon unitary
transition matrix: $ |\psi_{out}\rangle = U |\psi_{in}\rangle $, where (in the $|A\rangle ,|B\rangle ,|C\rangle $ basis) \begin{equation}U=-{i\over 3}\left(
\begin{array} {ccc} 1 & -2 & -2 \\ -2 & 1 & -2
\\ -2 & -2 & 1\end{array}\right) .\label{Umatrix}\end{equation} Up to overall phase, $U$ gives the reflection and transmission amplitudes of \cite{fh1,fh2,fh3}.

This matrix can also be found from more general unitarity and symmetry arguments, as follows. Assume a photon is sent into port $A$. Due to reflection symmetry
about the input direction, the exit amplitudes at the other two ports should be equal. Therefore, for  $|\psi_{in}\rangle =|A\rangle$, the output must be of the
form
\begin{equation}|\psi_{out}\rangle =  a|A\rangle +b\left( |B\rangle +|C\rangle \right) ,   \end{equation} for complex amplitudes $a$ and $b$.
Repeating the argument for inputs at ports $B$ and $C$, that the system's transition matrix $V$ must be of form \begin{equation}V=\left(
\begin{array}{ccc} a & b & b\\ b & a & b \\ b & b& a \end{array}\right) .\end{equation} The matrix must be unitary, so that $V\cdot V^\dagger =I$; the diagonal
and off-diagonal entries of this condition imply two constraints:
\begin{equation}|a|^2 +2|b|^2= 1,\qquad  2\mbox{Re}\left( ab^\ast\right) +|b|^2 =0.\end{equation}
Defining $a=\alpha e^{i\phi_a}$, $b=\beta e^{i\phi_b}$, and $\phi =\phi_b-\phi_a$, then solving these unitarity constraints gives:
\begin{equation}\alpha =\sqrt{ 1\over {1+8\cos^2\phi}}, \qquad \beta =-2\sqrt{ {\cos^2\phi}\over {1+8\cos^2\phi}}.\end{equation}
So the most general form of $V$ is:
\begin{equation}V={{e^{i\phi_a}}\over \sqrt{1+8\cos^2\phi}} \left( \begin{array}{ccc} 1 & -2\cos\phi & -2\cos\phi\\ -2\cos\phi & 1 & -2\cos\phi\\
-2\cos\phi & -2\cos\phi & 1\end{array}\right) .\end{equation} In the special case that $\phi=0$, this becomes
\begin{equation}V(\phi_a)={1\over 3}e^{i\phi_a} \left( \begin{array}{ccc} 1 & -2 & -2\\ -2 & 1 & -2\\ -2 & -2 & 1\end{array}\right) . \end{equation}
For $\phi_a=-{\pi\over 2}$ this gives back the matrix $U$ of Eq. \ref{Umatrix}.

It should be pointed out that, up to an overall phase, Eq. \ref{Umatrix} is the same as the well-known Grover coin for three-dimensional coined quantum walks. Similarly, the unbiased $n$-ports with $n>3$ that will be discussed briefly in Section \ref{generalsection}  correspond, when appropriate choices are made for the phases at the mirrors, to higher-dimensional Grover coins. Therefore, the unbiased $n$-ports can also be viewed as particularly simple linear-optical realizations of $n$-dimensional Grover coins. Note however that the unbiased $n$-ports are more general than this: by changing the phases and reflectances other coins can be implemented as well.

\section{Action on Bell states}\label{Bellsection} For two-photon entangled states, coherence requirements are less stringent than for single-photon states,
since it is primarily the much longer coherence time of the pump that is relevant, rather than the coherence times of the signal and idler. Further, spectral
filtering can stretch out the coherence time and remove distinguishability due to arrival time within each pair \cite{pan}. So it is natural to consider action
on two-photon states. Advantage can be taken of this in order to construct quantum gates in which the input and output qubits are not defined by identities of
individual photons, but rather by the symmetries shared by pairs of ingoing and outgoing two-photon states.

Applying appropriate control states, the multiport can convert any input Bell state into any output Bell state. The states considered here are distributed over
two ports, for example, $|\Psi^\pm\rangle_{AB}={1\over \sqrt{2}}\left( |H\rangle_A |V\rangle_B \pm |V\rangle_A |H\rangle_B \right) $ and
$|\Phi^\pm\rangle_{AB}={1\over \sqrt{2}}\left( |H\rangle_A |H\rangle_B \pm |V\rangle_A |V\rangle_B \right) $ at $A$ and $B$.  $H$ and $V$ denote horizontal and
vertical polarization.  Similar states are defined at other pairs of ports. Although mirror units interchange horizontal and vertical polarizations, every path
through the system encounters an even number of such units, leaving polarizations unchanged.

\begin{figure}
\centering
\includegraphics[totalheight=1.8in]{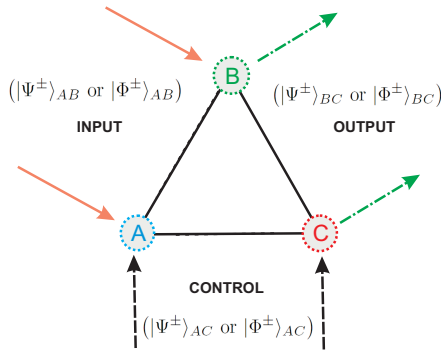}
\caption{(Color online.) Two-photon input enters ports $A/B$, with control state entering $A/C$. Output exits ports $B/C$. }\label{inoutfig}
\end{figure}

Consider creating two Bell states (by the procedure used in \cite{pan}, for example): a target state and a control state
(Fig. \ref{inoutfig}). These are input at ports $AB$ (input) and $AC$ (control). They are coupled into fibers, and the portions being fed into the same port are
merged using integrated Y couplers. The output, which can be separated from the input by an optical circulator, will be an entangled state shared by $B$ and $C$. Consider
states with one photon exiting at $B$ and one at $C$. (If multiple gates are concatenated, states with two photons exiting the same port need to be rejected.
Methods for doing this are known \cite{barrett}.) There will then be two photons exiting $A$. Acceptance of $BC$ output states is conditioned upon some outcome
at $A$. Choice of conditioning outcome provides control over the gate action. Consider two possibilities: either condition upon detecting two
photons of opposite polarization ($o$) or of same polarization ($s$) at $A$. It is only necessary to determine whether or not the polarizations are the same, not
to determine what the polarizations are (similar to the the probabalistic gates of \cite{pjf1,pjf2,pjf3,pjf4}).

Computing outputs for any input and control, and for either heralding condition, is straightforward (sample calculations are given in Appendix B), and result in
Table \ref{fulltable}. We see that the multiport acts as a universal processor on Bell states; by appropriate choice of control and heralding condition any input
state can be converted to any desired output state.


\begin{table}
\caption{Action on Bell states. The input is at $AB$, and control at $AC$, with output at $BC$ conditioned, respectively, on detecting the same ($s$) or opposite
($o$) polarizations at $A$.}
\begin{tabular}{|c|c|c|c| c |c |c|c|c|c|} \hline Input & Cont. & Out(s)  & Out(o) & & & Input & Cont. & Out(s)  & Out(o)\\ \hline
$\Psi^+$ & $\Psi^+$ & $\Phi^+$ & $\Psi^+$ & & & $\Psi^+$ & $\Phi^+$ & $\Psi^+$ & $\Phi^+$\\
$\Psi^+$ & $\Psi^-$ & $\Phi^-$ & $\Psi^-$ & & & $\Psi^+$ & $\Phi^-$ & $\Psi^-$ & $\Phi^-$\\
$\Psi^-$ & $\Psi^+$ & $\Phi^-$ & $\Psi^-$ & & & $\Psi^-$ & $\Phi^+$ & $\Psi^-$ & $\Phi^-$\\
$\Psi^-$ & $\Psi^-$ & $\Phi^+$ & $\Psi^+$ & & & $\Psi^-$ & $\Phi^-$ & $\Psi^+$ & $\Phi^+$\\ \hline
$\Phi^+$ & $\Phi^+$ & $\Phi^+$ & $\Psi^+$ & & & $\Phi^+$ & $\Psi^+$ & $\Psi^+$ & $\Phi^+$\\
$\Phi^+$ & $\Phi^-$ & $\Phi^-$ & $\Psi^-$ & & & $\Phi^+$ & $\Psi^-$ & $\Psi^-$ & $\Phi^-$\\
$\Phi^-$ & $\Phi^+$ & $\Phi^-$ & $\Psi^-$ & & & $\Phi^-$ & $\Psi^+$ & $\Psi^-$ & $\Phi^-$\\
$\Phi^-$ & $\Phi^-$ & $\Phi^+$ & $\Psi^+$ & & & $\Phi^-$ & $\Psi^-$ & $\Psi^+$ & $\Phi^+$
\\  \hline \end{tabular}\label{fulltable}\end{table}

\section{Probabilistic Controlled Entangled-State Gates}\label{gatesection} Taking subsets of Table \ref{fulltable}, we may implement probabalistic quantum gates. For example, take
$|\Psi^+\rangle_{AB}$ as input. Table \ref{psiplustable} shows how by varying just the control state and holding the heralding condition fixed, any desired state
can be selected from the possible output states. The same can be done using any other Bell input state.

\begin{table}
\caption{$\Psi^+$ can be converted to any desired output Bell state. Here the $s$ condition is assumed.}
\begin{tabular}{|c|c|c|} \hline Input & Control State &  Output\\ \hline
$\Psi^+$ & $\Psi^+$  & $\Phi^+$\\
$\Psi^+$ & $\Psi^-$  & $\Phi^-$\\
$\Psi^+$ & $\Phi^+$  & $\Psi^+$\\
$\Psi^+$ & $\Phi^-$  & $\Psi^-$\\ \hline
\end{tabular}\label{psiplustable}\end{table}

As one example of symmetry-based Bell state processing, the multiport implements probabilistic CNOT gates for entangled states. Take $|\Psi^\pm\rangle$ states as
input and control, but  $|\Phi^\pm\rangle$ states as output, with $+$ states of either type corresponding to $|0\rangle$ and $-$ states to $|1\rangle$ in all
cases. (So the bit is determined by the polarization-interchange symmetry, not by the particular state that happens to be carrying the symmetry.) Attention is
then restricted to the top left quadrant of Table I with condition $s$, yielding a CNOT truth table. For multiple processing steps through multiple gates, the
roles of $|\Psi^\pm\rangle$ and $|\Phi^\pm\rangle$ flip on alternate clock steps: output $|\Phi^\pm\rangle$ states at one step would become input for the next
step, with $|\Psi^\pm\rangle$ states then being output at that next step. On alternate steps, the bottom left quadrant of Table I would be used. Success
probability for this \emph{four-photon} gate is about $5\%$, not far below the lower end of the success probability range for proposed \emph{two-photon}
probabilistic gates ( $\sim {1\over 6}$ to ${1\over {2}}$) \cite{pjf1,pjf2,pjf3,pjf4,klm,koashi,uskov}.
Instead of encoding qubits into the state symmetry, the Bell states can also be viewed as qudits on a four-dimensional Hilbert space. The multiport then would
act as a four-photon, \emph{two-qudit} gate.


\section{Group structure}\label{groupsection} 
\begin{figure}
\centering
\includegraphics[width=2.4in]{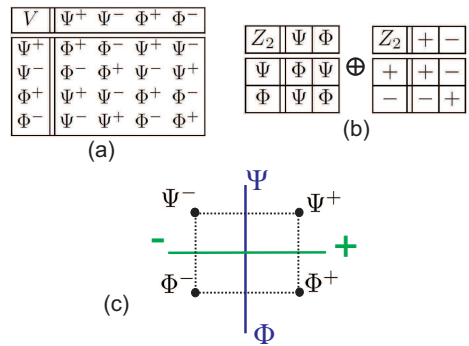}
\caption{(Color online.) The multiport produces the Abelian 4-group multiplication table (a). (b) Cyclic $Z_2$ subgroups describe reflections about axes
shown in (c).}
\label{grouptable}\end{figure}

The device turns two ingoing states into one output state. When the $s$ condition is used, this imposes an Abelian group structure on
the states. (For $o$ condition, the same structure appears, with $\Phi^\pm $ and $\Psi^\pm$ interchanged.) The resulting group multiplication table of Fig.
\ref{grouptable}(a) has rows and columns labeling input and control states; output occupies the interior. This is the Klein $4$-group \cite{armstrong}, $V$,
which is a direct sum of two cyclic groups, $V=Z_2\oplus Z_2$ (Fig. \ref{grouptable}(b)). The action of this group is not difficult to understand: consider the
two-dimensional space with one axis labeled by $+$ and $-$ (right and left), the other by $\Psi$ and $\Phi$ (up and down). Bell states are then at corners of a
rectangle, and the $Z_2$ groups reflect the axes (Fig. \ref{grouptable}(c)). If input states are restricted to $\left\{|\Psi^+\rangle,|\Phi^+\rangle\right\}$,
then a single $Z_2$ group arises.

This raises interesting possibilities. Post-selection introduces effective "interactions" between photons with states acting on each other via mathematical
groups. The triangular device is suited for this, with two sides to use for the input of the two ingoing states, and the third side for the group product output. Because all vertices are taken to be identical, it is clear that only Abelian groups can occur here: interchanging which state enters which side can have no effect on the outcome. However, when using different complex reflectances
at different vertices, the symmetry of the device is broken; as a result, changing the order of multiplication of the states (which state is applied to which side of the triangle) will now affect the outcome. In this latter case any groups that arise would be expected to be non-Abelian.
In addition, it remains to be investigated whether larger groups could be obtained from
larger sets of input states and from multiports with more than three vertices. In any case, the
ability to engineer groups of states opens new avenues of investigation on optically-implemented group representations for quantum information
processing, such as group-based strategies for quantum cryptography or group-based security mechanisms for quantum key distribution.

As one example of why a group approach may be useful, consider information processing with $d$-state logic. Imagine that possible input and output states form a
faithful representation in Hilbert space of a group $G$ of order $d$. The full range of operations possible on this set of states (the transformation of any of the $d$ inputs
to any of the $d$ outputs) would require a total of $d$ operations. But suppose that the group is of rank $r$, with a set of generators, $g_1,\dots, g_r$.
This means that any element $g\in G$ can be written in the form $g=g_1^{n_1}\cdot g_2^{n_2}\cdot \dots g_r^{n_r}$ for some appropriate set of integers $n_1,\dots
,n_r$. This is the case for the Klein group, for example, where the four-element group is generated by the two generators of the $Z_2$ subgroups. Similarly a set of states lying in a plane and related by a discrete rotation group of \emph{any} rank can always be spanned by
powers of just a single generator. Therefore $d$-state logic can be carried out on a group representation in Hilbert space with only
$r$ types of basic logic units. In the directionally-unbiased scheme presented here, different logic units would correspond to triangular (or more general polygonal) units with the vertex
parameters possibly set to different values or with different control states. In the case where $r$ is much smaller than $d$, then very high-order logic systems
can be implemented with a small number of different logic units. This increased parsimony is illustrated by the construction of Sections \ref{Bellsection} and
\ref{gatesection}, where four different operations on a given target state are carried out by a \emph{single} triangular controlled logic unit.

As mentioned in the introduction, new possibilities may be raised for quantum cryptographic schemes as well. For example, there is also a connection between the approach of
this paper and a group theory-related quantum key distribution protocol that has already been proposed. This is the quantum enigma scheme \cite{lloyd,guha},
which is based on quantum data locking: a small amount of transmitted information can unlock a much large trove of data to a user, while keeping the data secure
from unapproved agents. This is done by using a set of $N\times N$ unitary matrices (i.e. elements of the group $U(N)$) to rotate between a collection of
mutually unbiased bases. A randomly chosen element of this set is applied to an initial state. An eavesdropper cannot access the information in the transmitted
state without knowing what unitary transformation is needed to rotate it back to its original basis. Rather than sending the entire unitary matrix, only a
discrete label identifying the matrix needs to be securely transmitted. The approach of the current paper constructs discrete group transformations, which can be
embedded into the continuous groups $U(N)$, and so can be viewed as a physical implementation of the quantum
enigma transformations for low $N$. Going back again to the example of the Klein group, the four Bell states are related by a discrete subgroup of the unitary group $U(2)$ acting on the two-dimensional complex space spanned by the basis $|H\rangle$ and $|V\rangle$. To be useful for the enigma procedure, the current setup must be generalized by using more complex input and output states, in order to allow larger subsets of higher $U(N)$ groups to appear.


\section{Generalizations and further directions}\label{generalsection}  
\begin{figure}
\centering
\includegraphics[totalheight=1.8in]{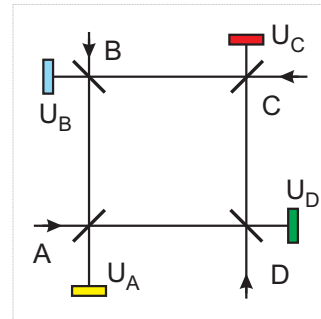}
\caption{(Color online.) Directionally-unbiased four-port. Generalizations to any number of ports $\ge 3$ are possible. }\label{fourfig}
\end{figure}

The triangular geometry can be replaced by any regular $n$-sided polygon, such as the $4$-port in Fig. \ref{fourfig}. This enlarges the symmetry
group of the device, and allows imposition of other, more complex group structures on the input and output states. That would, in turn, expand the potential
capacity for information processing applications. Using arbitrary polygonal units allows implementation of scattering experiments on undirected graphs with nodes
of arbitrary valence.

For $n$-ports with $n>3$, if $d$ is not an integer multiple of the wavelength, then paths between ports separated by different distances will gain different phase shifts and different amplitude losses (due to the different number of intervening beam splitters crossed). The phase shifts can cause possible interference effects within the multiport. As a result, the off-diagonal terms in $U$ are no longer all equal. This effect will be unavoidable for finite pulses with a nonzero frequency spread. However, if $d$ is an integer multiple of the central wavelength of the pulse, then the effect will again be small if the coherence time is long, or in other words, if the spread of wavelengths away from the central value is small. Even for long $\tau_{coh}$ this will become eventually become significant if the multiport is made sufficiently large (in the sense of having a large number of input/output ports), but if the details of the pulse are known, then the effect on $U$ can be calculated. The overall symmetries of the multiport under reflection and rotation will still be maintained, but the output will depend on the distance between the input and output ports, as reflected by the change in the off-diagonal entries of $U$.

In this paper a single triangular unit has been considered in isolation. When connected in networks, new possibilities arise. For example, quantum walks
on lattices with different symmetries and with vertices having easily controllable properties can be constructed in a simple manner. The fact that all of the
elements of the basic polygonal multiport can be put onto a single optical chip means that large networks can be be readily constructed with fewer alignment and
stability problems than would occur using tabletop arrangements of discrete beam splitters and mirrors. Different polygons can be joined together in the same
network (for example triangular, square, and pentagonal multiports) in order to experimentally study more complex systems or behavioral changes in the system due
to transition from one symmetry type to another. Scattering systems on undirected graphs can then be studied experimentally, including graphs with different valences
at different nodes.

Variation of the system's parameters can also be introduced. Reflectances and mirror unit phase shifts may differ at each corner, allowing tailoring of a range
of output qutrits from the same input state by parameter tuning. In two-dimensional networks of such units, this parameter tuning allows introduction of
controlled spatial bias into quantum walks, which is useful for algorithmic applications \cite{lu}.
Further, parameters can be varied
while a walk is in progress, allowing investigations into time-dependent quantum walks.

\section{Conclusions}\label{concludesection}

In this paper, linear optical unbiased multiports were introduced. Symmetry arguments and path tracing show that, despite its simplicity, this system has a
number of unusual properties and applications. These devices allow simple implementations of optical scattering experiments on graphs, information processing
directly on Bell states or other more complicated state sets, and information processing on optically-implemented group structures. Some tentative possible
directions have been raised for applications of a group-based approach to optical cryptography and information processing.

It should also be noted that the current directionally-unbiased framework differs in several important ways from all linear-optical information
processing schemes that fall into the Knill-Laflamme-Milburn (KLM) \cite{klm} framework. For example, KLM-based schemes always consider only output states with
at most a single photon in each output mode. Those output states with multiple photons in the same mode are considered to be signals that the desired operation
has failed, and so they are discarded. In contrast, in the scheme described in this paper constructive use is made of the possibility that multiple photons may
leave the device in the same mode. Similarly, in KLM-type setups, the input and output ports are always distinct from each other, with the photons always
traveling in a single direction through the network. The introduction of directionally unbiased propagation in our scheme allows the merging of input and output
ports, which can greatly decrease the complexity of an optical network; for example, the number of beam splitters in general scales quadratically in $N$ for KLM
networks, where $N$ is the number of photons, or equivalently, the number of input or output ports. In the present case, the unbiased $N$-port setup can process $N$ photons through $N$ input/output ports with just $N$ beam splitters
and $N$ mirrors; it therefore scales only linearly with $N$. For large $N$, this drop in scaling from quadratic to linear is an enormous savings of resources.

A rich range of further possibilities remain to be explored, such as optical implementation of three-state logic using qutrits on the triangular multiports, or
of higher-state logic on larger polygonal multiports.

\section*{Acknowledgements.} This research was supported by the DARPA QUINESS program through US Army Research Office award W31P4Q-12-1-0015, by the National
Science Foundation under Grant No. ECCS-1309209, and by Northrop Grumman Aerospace Systems.

\appendix
\section{Paths and amplitudes through the device}

Here, the paths of length $N\le 10$ that start at $A$ and exit the system are tabulated, along with their amplitudes. A few other properties of the transition
matrix $U$ are also given for completeness.

\begin{figure*}[h!]
\begin{center}
\subfigure[]{
\includegraphics[height=3.0in]{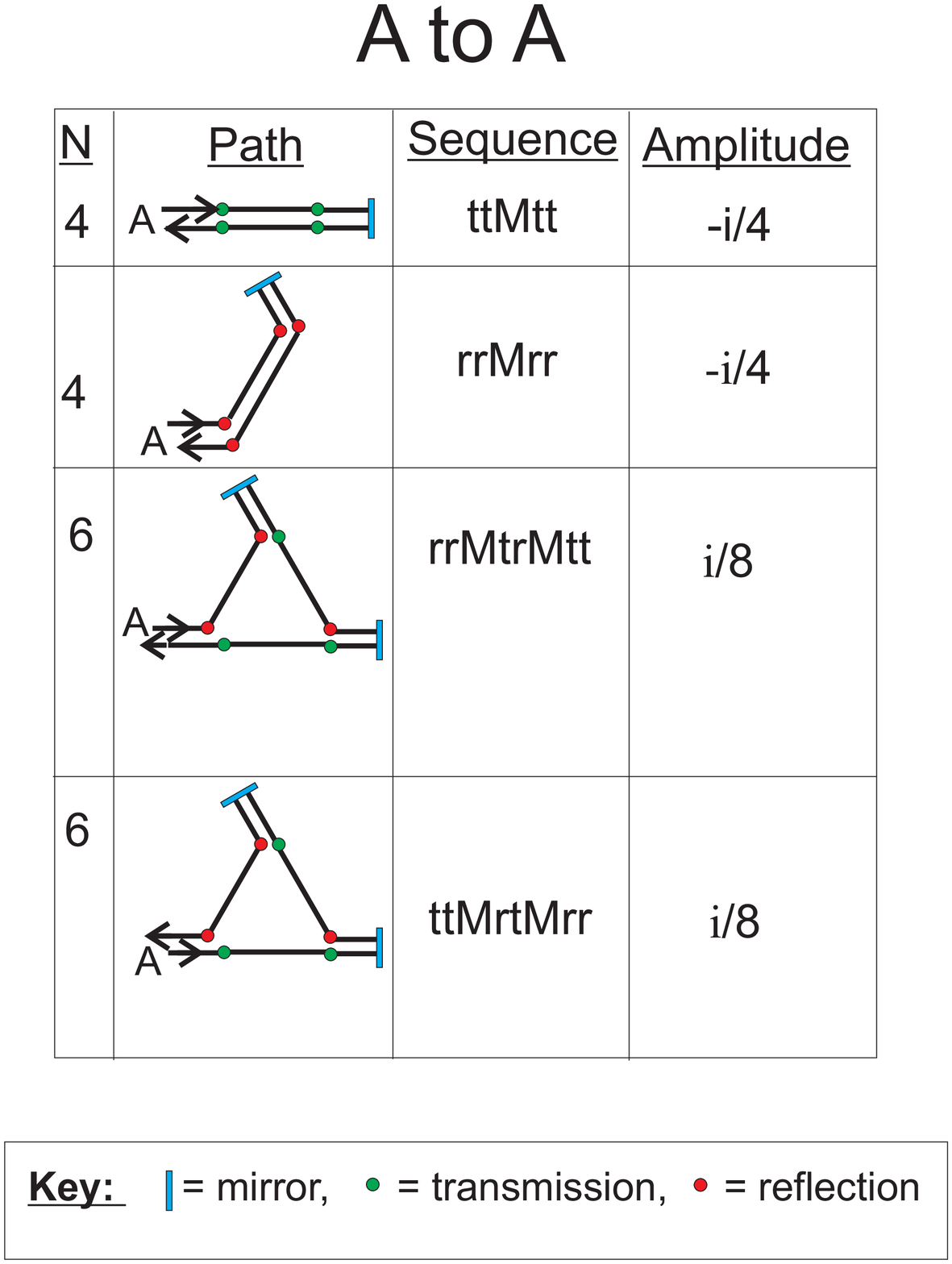}}
\subfigure[]{
\includegraphics[height=3.0in]{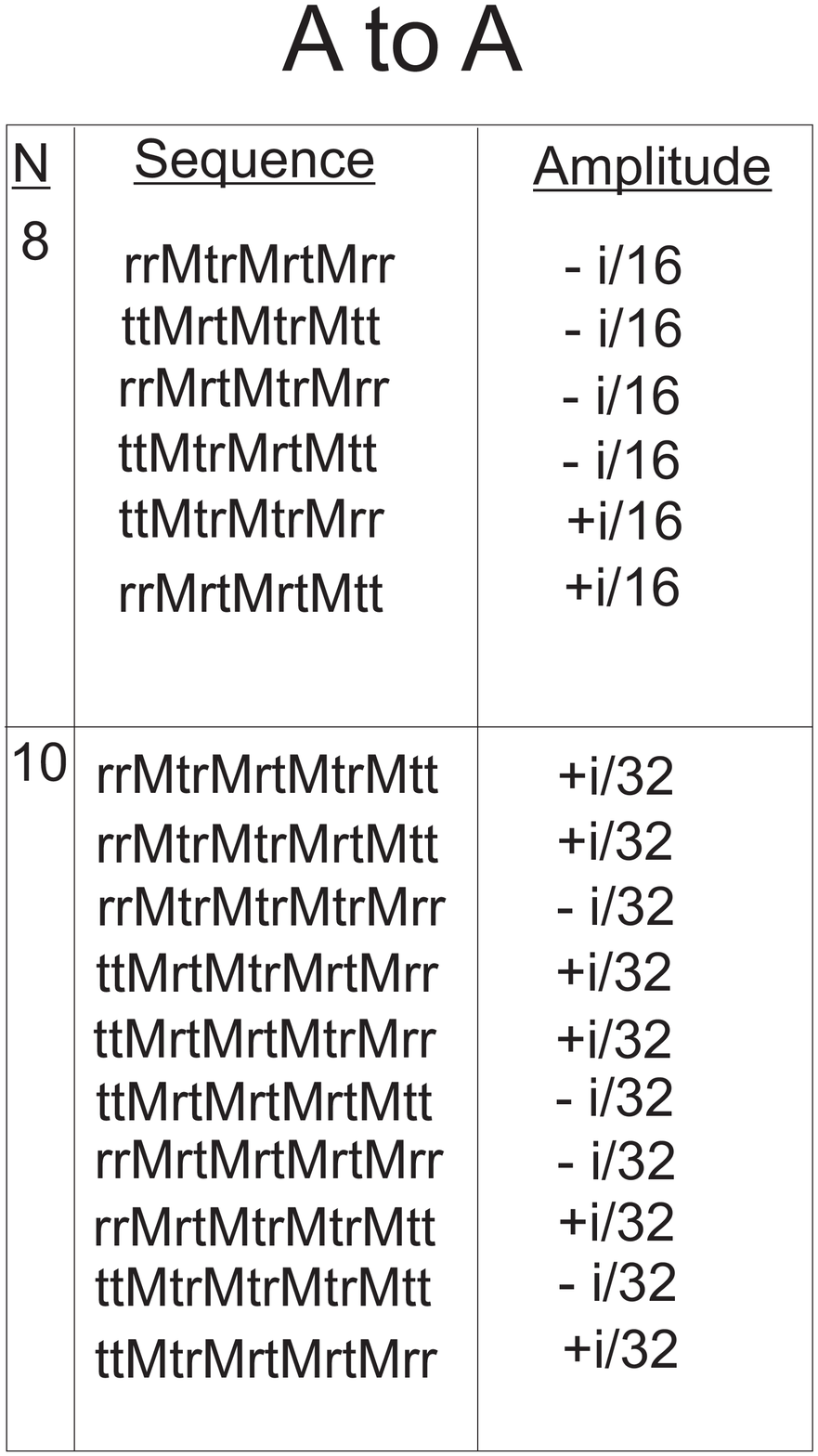}}
\caption{(Color online.) The paths of length up to $N=10$ that enter port $A$ and then exit at the same port.}\label{AtoA}
\end{center}
\end{figure*}

\begin{figure*}[h!]
\begin{center}
\subfigure[]{
\includegraphics[height=3.0in]{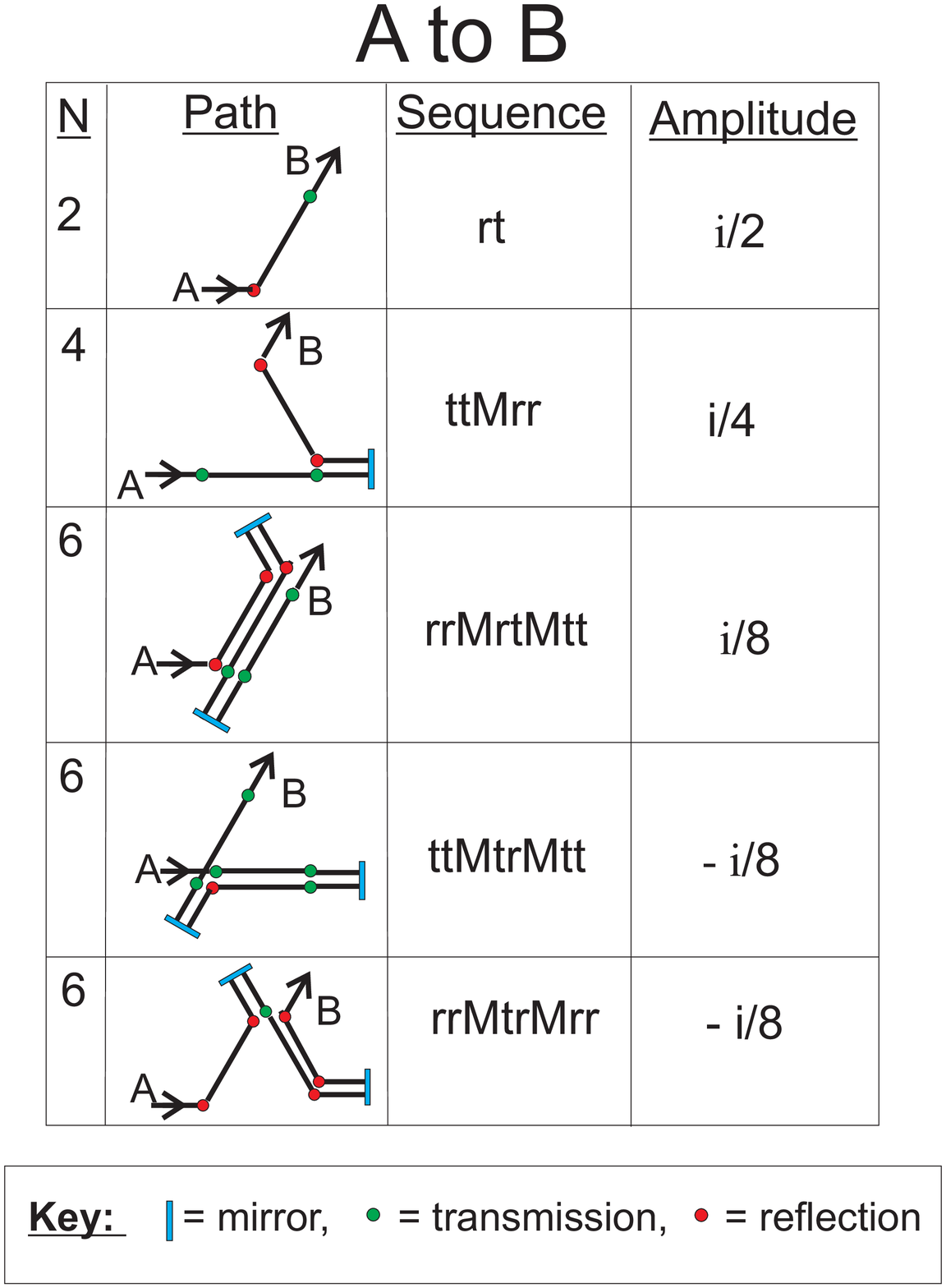}}
\subfigure[]{
\includegraphics[height=3.0in]{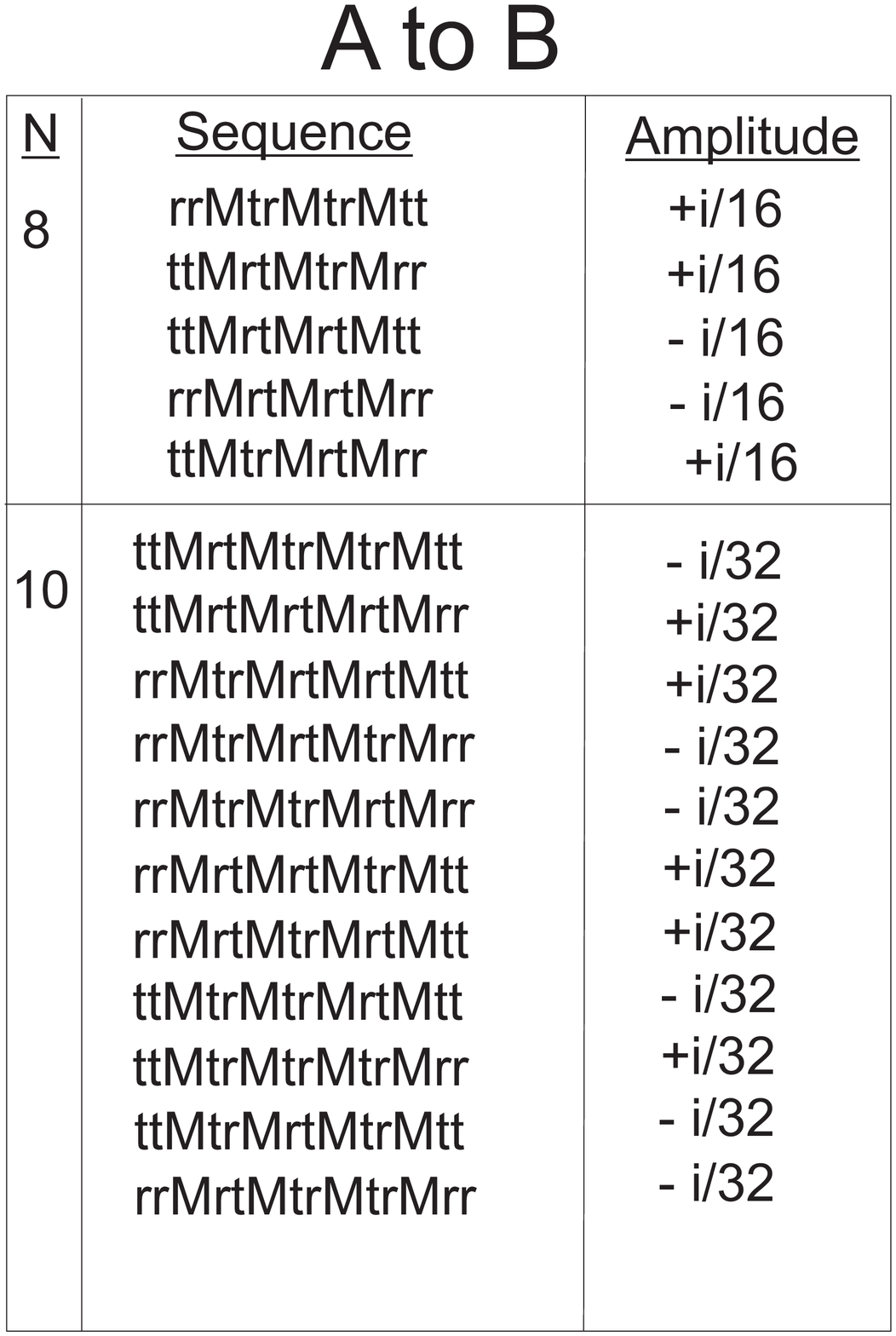}}
\caption{(Color online.) The paths of length up to $N=10$ that enter port $A$ and then exit at the port $B$.}\label{AtoB}
\end{center}
\end{figure*}

\begin{figure*}[h!]
\begin{center}
\subfigure[]{
\includegraphics[height=3.0in]{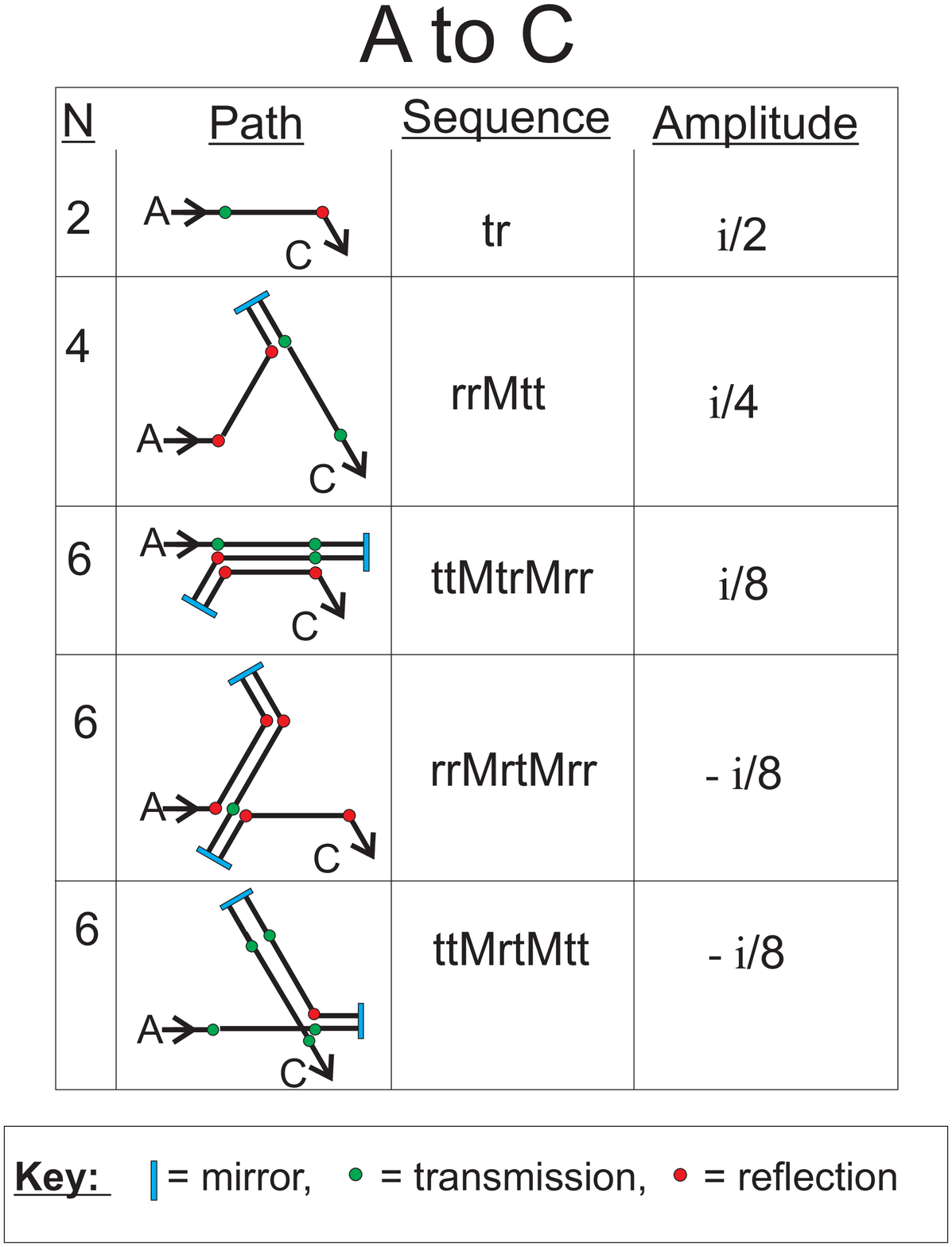}}
\subfigure[]{
\includegraphics[height=3.0in]{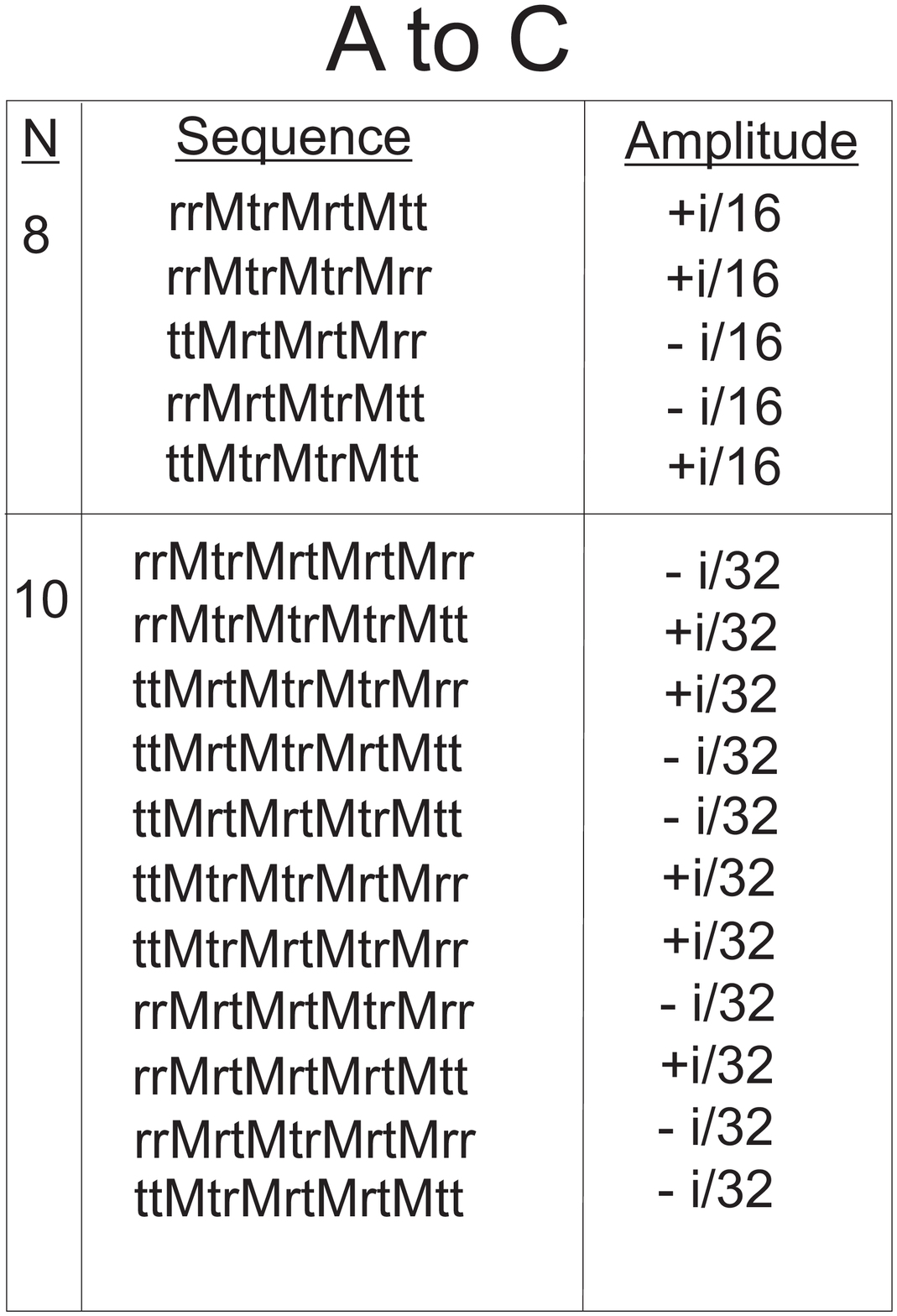}}
\caption{(Color online.) The paths of length up to $N=8$ that enter port $A$ and then exit at the port $C$.}\label{AtoC}
\end{center}
\end{figure*}

The possible paths are given in the tables of Figs. \ref{AtoA}-\ref{AtoC}. In each line, the path is given in symbolic form, with $r$, $t$, and $M$ respectively
representing reflection or transmission at a beam splitter and reflection at a mirror. Drawings of the paths after $N=6$ are not included because they start
becoming too cumbersome. The sequences of reflections and transmissions are ordered from left (first one applied) to right (last). Although it is assumed that
the input is at $A$, there is no loss of generality: the paths for input at the other ports can immediately be obtained from these by cyclic permutation. $50/50$
beam splitters are assumed, with factors of $i$ for beam splitter reflections and a total reflection coefficient $-i$ at each mirror unit.

Summing the entire infinite series  of paths leads to the transition matrix $U$ given in the main text. $U$ has one eigenvector with eigenvalue $+i$, given by \begin{equation}{1\over \sqrt{3}}\left(\begin{array}{c} 1 \\ 1\\ 1\end{array} \right) =  {1\over \sqrt{3}} \left( |A\rangle +|B\rangle +|C\rangle\right) ,\\
\end{equation} where $|A\rangle$, $|B\rangle$, and $|C\rangle$ are defined in the main text. The remaining two eigenvectors are degenerate, with eigenvalue $-i$, and
may be taken to be any two of the three linearly-dependent vectors \begin{eqnarray} {1\over \sqrt{2}}\left(\begin{array}{c} 1 \\ -1\\ 0\end{array} \right)
&=& {1\over \sqrt{2}} \left( |A\rangle -|B\rangle \right)\\
{1\over \sqrt{2}}\left(\begin{array}{c} 1 \\ 0\\ -1\end{array} \right)&=& {1\over \sqrt{2}} \left( |A\rangle -|C\rangle \right)\\ {1\over \sqrt{2}}
\left(\begin{array}{c} 0 \\ 1\\ -1\end{array} \right) &=& {1\over \sqrt{2}} \left( |B\rangle -|C\rangle \right) .\end{eqnarray} The eigenstates are therefore
those that are either antisymmetric about any pair of ports or completely symmetric about all three ports. It may also be noted that $U^2=-I$, where $I$ is the
$3\times 3$ identity matrix.

The probability of exiting at a given time is given by adding the amplitudes of the indistinguishable paths that exit at the same port, then adding the
probabilities of the different ports. For even $N$ ($N>2)$, the instantaneous exit probability at time $t=NT$ is $6/N^2$, so that the cumulative probability of
exiting by time $NT$ is $${1\over 2}+{3\over 2}\sum_{n=2}^{N/2} {1\over {n^2}}.$$ Exit probabilities at odd $N$ vanish.

\section{Sample calculation for processing of Bell states}

Here, an example is given of the calculations done to compile Table II. Suppose that the input state is $|\Psi^+\rangle_{AB}$ and the control state is
$|\Psi^+\rangle_{AC}$. Then the action of the multiport is
\begin{equation}|\Psi^+\rangle_{AB}\otimes |\Psi^+\rangle_{AC} \to (U\otimes U)|\Psi^+\rangle_{AB}\otimes (U\otimes U)|\Psi^+\rangle_{AC},\end{equation} where a few lines of algebra
starting from Eqs. (5) and (6) of the main text give
\begin{eqnarray} & & (U\otimes U)|\Psi^+\rangle_{AB}\\ & &  = -{1\over 9} \left\{ 2\sqrt{2}\Big( |H\rangle_A|V\rangle_A  +|H\rangle_B|V\rangle_B
-2|H\rangle_C|V\rangle_C\Big)\right. \nonumber \\ & & \qquad\qquad  \left. -5|\Psi^+\rangle_{AB} -2\Big( |\Psi^+\rangle_{AC}+|\Psi^+\rangle_{BC}\Big) \right\} ,\nonumber
\end{eqnarray} with a similar result for $U|\Psi^+\rangle_{AB}$.
Taking the product $\left(U\otimes U|\Psi^+\rangle_{AB}\right)\otimes \left( U\otimes U|\Psi^+\rangle_{AC}\right) $ gives \begin{equation}{1\over {81}} \left\{
-8\sqrt{2}|H\rangle_A|V\rangle_A |\Psi^+\rangle_{BC}  +29|\Psi^+\rangle_{AC}|\Psi^+\rangle_{AB} +\dots \right\} ,
\end{equation} where the dropped terms are those that do \emph{not} have a single photon at $B$ and a single photon at $C$. Note that \begin{eqnarray}
& & |\Psi^+\rangle_{AB}|\Psi^+\rangle_{AC} = {1\over 2} \left\{ \sqrt{2} |2H\rangle_A|V\rangle_B|V\rangle_C \right. \\ & & \qquad + \sqrt{2}
|2V\rangle_A|H\rangle_B|H\rangle_C \nonumber\\ & & \qquad \left.  + |H\rangle_A|V\rangle_A\Big( |V\rangle_B|H\rangle_C+|H\rangle_B|V\rangle_C\Big) \right\}
,\nonumber
\end{eqnarray} where, for example, $|2H\rangle_A$ is the state with two horizontally polarized photons at $A$. Using this, the outgoing state becomes
\begin{eqnarray} & & {1\over {81}} \left\{  13\sqrt{2} |H\rangle_A|V\rangle_A |\Psi^+\rangle_{BC} \right. \\ & & \quad \left.
+{{29}\over \sqrt{2}} \Big[ |2H\rangle_A|V\rangle_B|V\rangle_C +|2V\rangle_A|H\rangle_B|H\rangle_C \Big] +\dots \right\}.\nonumber \end{eqnarray} By projecting
onto the part with opposite polarizations at $A$  (the first term), the state $|\Psi^+\rangle_{BC}$ is picked out, while projecting onto the portion with same
polarizations at $A$ (second and third terms) picks out the state \begin{equation}{1\over \sqrt{2}} \Big[ |V\rangle_B|V\rangle_C + |H\rangle_B|H\rangle_C \Big]
=|\Psi^+\rangle_{BC}.\end{equation} Repeating the same procedure for all possible products of input and control states then fills out the entries in Table II.

\vfill


\begin{thebibliography}{99}

\bibitem{falci} G. Falci, E. Paladino, Int. J. Quant. Inf.  \textbf{12}, 1430003 (2014)

\bibitem{blackburn} S. R. Blackburn, C. Carlos, C. Mullan, in \emph{Groups St Andrews 2009 in Bath}, eds. C. M. Campbell, et. al. (Cambridge University Press, Cambridge, 2011), pp. 133-149.

\bibitem{vasco} M. I. Gonz\'{a}lez Vasco, R. Steinwandt, \emph{Group Theoretic Cryptography} (Chapman and Hall/CRC, Boca Raton, 2015).

\bibitem{genovese} M. Genovese, Phys. Rep. \textbf{413}, 319 (2005)

\bibitem{aharanov} Y. Aharonov, L. Davidovich, N. Zagury, Phys. Rev. A \textbf{48}, 1687 (1993).

\bibitem{kempe} J. Kempe, Cont. Phys. {\bf 44}, 307 (2003).

\bibitem{andraca} S. E. Venegas-Andraca, Quant. Inf. Process. \textbf{11}, 1015 (2012).

\bibitem{portugal} R. Portugal, \emph{Quantum Walks and Search Algorithms} (Springer, Berlin, 2013).

\bibitem{childs} A. M. Childs, 
Phys. Rev. Lett {\bf 102}, 180501 (2009).

\bibitem{fh1} E. Feldman, M. Hillery, Phys. Lett. A \textbf{324}, 277�281 (2004).

\bibitem{fh2} E. Feldman, M. Hillery, Cont. Math. \textbf{381}, 71 (2005).

\bibitem{fh3} E. Feldman, M. Hillery, J. Phys. A: Math. Theor. \textbf{40}, 11343 (2007).

%
%
%

\bibitem{saleh} B. E. A. Saleh, M. C. Teich, {\it Fundamentals of Photonics, 2nd ed.} (Wiley-Interscience, Hoboken, 2007).

\bibitem{pan} J. W. Pan, M. Daniell, S. Gasparoni, G. Weihs, A. Zeilinger, Phys. Rev. Lett. \textbf{86}, 4435, (2001).

\bibitem{barrett} S. D. Barrett, P. Kok, K. Nemoto, R. G. Beausoleil, W. J. Munro, T. P. Spiller, Phys. Rev. A {\bf 71}, 060302(R) (2005).

\bibitem{pjf1} T. B. Pittman, B. C. Jacobs, and J. D. Franson, Phys. Rev. A \textbf{64}, 062311 (2001).

\bibitem{pjf2} T. B. Pittman, B. C. Jacobs, and J. D. Franson, \emph{ Phys. Rev. Lett.} \textbf{88},  257902 (2002).

\bibitem{pjf3} T. B. Pittman, M. J. Fitch, B. C. Jacobs, and J. D. Franson, Phys. Rev. A \textbf{68}, 032316 (2003).

\bibitem{pjf4} T. B. Pittman, B. C. Jacobs, and J. D. Franson, Phys. Rev. A \textbf{71}, 032307 (2005).

\bibitem{klm} E. Knill, R. Laflamme, and G. J. Milburn, Nature (London)
\textbf{409}, 46 (2001).

\bibitem{koashi} M. Koashi, T. Yamamoto, and N. Imoto, Phys. Rev. A \textbf{63},
030301 (2001).

\bibitem{uskov} D. B. Uskov, P. M. Alsing, M. L. Fanto, L. Kaplan, A. M. Smith,  arXiv:1306.4062 [quant-ph] (2013)

\bibitem{armstrong} M. A. Armstrong, \emph{Groups and Symmetry} (Springer, Berlin, 1988).

\bibitem{lloyd} S. Lloyd, Quantum Engima Machines, arXiv:1307.0380 (2013)

\bibitem{guha} S. Guha, P. Hayden, H. Krovi, S. Lloyd, C. Lupo, J. H. Shapiro, M. Takeoka, and M.
    M. Wilde, Phys. Rev. X \textbf{4}, 011016 (2014)

\bibitem{lu} D. Lu, J. D. Biamonte, J. Li, H. Li, T. H. Johnson, V. Bergholm, M. Faccin, Z. Zimbor\'as, R. Laflamme, J. Baugh, S. Lloyd, Phys. Rev. A
    \textbf{93}, 042302 (2016).
\end{thebibliography}
\end{document}